\documentclass[pra,twocolumn,superscriptaddress,preprintnumbers,amsmath,amssymb]{revtex4}
\usepackage{}
\usepackage{}

\usepackage{color}
\usepackage[colorlinks=true]{hyperref}
\hypersetup{colorlinks=true,linkcolor=blue,citecolor=blue,urlcolor=blue}
\usepackage{bm}
\usepackage{bbm}
\usepackage{amssymb}
\usepackage{amsfonts}
\usepackage{epsfig,graphicx}
\usepackage{amstext}
\usepackage{amsmath}
\usepackage{graphicx}
\usepackage{times}
\usepackage{txfonts}
\usepackage{dcolumn}
\usepackage{dsfont}
\usepackage{braket}

\newcommand{\tr}{\mathrm{tr}}

\begin{document}

\title{Enhancing the charging performance of an atomic quantum battery}

\author{Ming-Liang Hu}
\email{mingliang0301@163.com}
\address{School of Science, Xi'an University of Posts and Telecommunications, Xi'an 710121, China}

\author{Ting Gao}
\address{School of Science, Xi'an University of Posts and Telecommunications, Xi'an 710121, China}

\author{Heng Fan}
\email{hfan@iphy.ac.cn}
\affiliation{Institute of Physics, Chinese Academy of Sciences, Beijing 100190, China}
\affiliation{School of Physical Sciences, University of Chinese Academy of Sciences, Beijing 100190, China}
\affiliation{Beijing Academy of Quantum Information Sciences, Beijing 100193, China}

\begin{abstract}
We study a quantum battery (QB) model composed of two atoms, where the charger and battery elements are coupled to a multimode vacuum field that serves as a mediator for energy transfer. Different figures of merit such as ergotropy, charging time, and charging efficiency are analyzed, putting emphasis on the role of various control parameters on the charging performance. It is found that there is a range of angle between the transition dipole moments and interatomic axis in which the QB can be charged. The optimal charging performance is achieved if the atomic dipole moments are perpendicular or parallel to the interatomic axis. The charging performance also improves with the decrease of the interatomic distance. Besides, the charged ergotropy can be enhanced by increasing the initial ergotropy of the charger and it is beneficial to charge the QB starting from a passive state.
\end{abstract}


\maketitle

\section{Introduction} \label{sec:1}
Quantum batteries (QBs) are small quantum devices that utilize quantum effects to achieve efficient energy storage and distribution \cite{EPL,collective1}. A central quantity characterizing the performance of a QB is ergotropy, which quantifies the maximal amount of energy that can be extracted from it via cyclic unitary operations in order to perform thermodynamic work \cite{EPL}. With the rapid developments in this field, it was realized that a good QB should have not only large internal energy and ergotropy, but also high charging rate, charging power, and energy conversion efficiency. In this regard, it was recently shown that the performance of a QB could be improved by using collective operations \cite{collective1,collective2,collective3,collective4,collective5,collective6,correlated} or by using quantum resources such as entanglement, discord, and coherence \cite{resource1,resource2,resource3, resource4,resource5,resource6,AVS}.

During recent years, various proposals for implementations of QBs have been reported, such as the cavity QBs \cite{collective4,resource2,cavity1,TCqb1,Dicke1,Dicke2,Dicke3,cavity2,cavity3}, the spin-chain QBs \cite{resource5, spinb1,spinb2, spin1,spin2,spin3,spin4,spin5,spin6}, the Sachdev-Ye-Kitaev QBs \cite{SYKb1,SYKb2}, and so on \cite{threelevelqb,oscillator0,oscillator1,oscillator2,oscillator3}. There are also several works for experimental implementations of these devices \cite{exp1,exp2,exp3,exp4,exp5}. One of the motivations for constructing these models is to see whether quantum effects, which play a central role in quantum information processing and quantum metrology, can also help to improve the capabilities of these energy-storing devices. If this is indeed the case, then other intriguing questions centering around this issue are how to formulate a quantum description of their working mechanism and how to understand their performance under various scenarios \cite{resource6,workext2,passive3,shihl2,luomx}.

Considering the realistic situations, the charger and battery elements are inevitably affected by the environments, thereby the dissipative charging of a QB has also been extensively studied, aimed at seeking schemes to mitigate the deterioration in the charging processes \cite{collective5,collective6, resource3,oscillator2,noiseqb1,noiseqb2,PRAppl,noiseqb3,noiseqb4,noiseqb5,noiseqb6,noiseqb7,squeezqb,anjh,weak}. For example, it was shown recently that the performance deterioration due to dissipation can be mitigated by scaling up the battery size \cite{collective6}, and a squeezed thermal reservoir can help to charge a QB efficiently \cite{squeezqb}. Moreover, some dissipative reservoirs can even be exploited as mediators for implementing wireless charging of a QB \cite{wireless1,wireless2,wireless3}. Other related investigations include general bounds for the power of collective charging \cite{bound0, bound1,bound2} and the collective advantage of a QB \cite{collective3,SYKb1,advan1,advan2,advan3, advan4,advan5,advan6,advan7}. Specifically, the collective operations on $n$ collective battery cells may extract more work than the sum of work extractable from each battery cell, the property of which has its toots in the principle that the product of two independent copies of a passive state may not always be passive \cite{collective2}. Many efforts have also been dedicated to improving the performance of a QB in terms of charging capacity, charging power, charging rate, and total stored energy \cite{collective3,chpf1,thermalization,chpf2,chpf3,chpf4,chpf6, chpf7, Shangc}.

In reality, it is possible to charge a QB starting from a general condition, that is, starting from the condition under which the charger energy is not maximal and the QB is not empty; in particular, it is appealing to transfer the residual charger energy (even it is not maximal) to a QB. However, most of the previous works considered only the case that the charger is in its excited state and the QB is in its ground state initially. As we know, the evolution trajectory of a system may be completely different starting from different initial conditions. Thereby, the initial conditions are also expected to be important in the charging process. In fact, a recent study has been carried out putting emphasis on the role of different preparation states on the performance of a QB driven by a time dependent classical force \cite{initial}.

In this work, we study a QB model composed of two atoms interacting with a multimode vacuum field. The first atom is treated as the charger and the second one as the QB, while the field plays the role of a mediator for energy exchange between them. The mutual interaction of the two atoms with the multimode vacuum field leads to atomic spontaneous emission, collective damping, and dipole-dipole interaction, the latter two of which depend on the angle  between the transition dipole moments and the interatomic axis as well as the separation between the two atoms. Here, we fully incorporate these factors and analyze in detail how they affect different figures of merit characterizing the performance of the atomic QB model initialized in different states.

The structure of this paper is organized as follows. In Sec. \ref{sec:2}, we present the QB model we considered and recall the preliminaries related to ergotropy. In Sec. \ref{sec:3}, we discuss in detail the charging performance of the proposed atomic QB under different initial conditions. Finally, we conclude this paper with a short summary in Sec. \ref{sec:4}.

\section{The charger-battery model} \label{sec:2}
We consider a QB model consisting of two two-level atoms. The lower level $|0\rangle$ and upper level $|1\rangle$ of them are separated by an energy gap $\hbar\omega_0$, where $\omega_0$ is the transition frequency. We treat the first atom as the charger and the second one as the QB, where the indirect interaction between them is mediated by their mutual interaction with a multimode vacuum field, which plays the role of a mediator for transferring energy from the charger to the QB. The master equation (in units of $\hbar$) governing the dynamics of this system reads \cite{ME0,ME1,ME2,ME3}:
\begin{equation}\label{eq2-1}
\begin{aligned}
\frac{\partial \rho}{\partial t}= \, & -i\omega_0 \sum_{i=1}^2 [S_i^z, \rho]- i\sum_{i\neq j}^2 \Omega_{ij} [S_i^{+}S_j^{-}, \rho] \\
                                     & -\frac{1}{2}\sum_{i,j=1}^2 \gamma_{ij} \big(S_i^{+}S_j^{-} \rho + \rho S_i^{+}S_j^{-}
                                       -2S_i^{-} \rho S_j^{+} \big),
\end{aligned}
\end{equation}
where $S_i^z$ is the energy operator of the $i$th atom and $S_i^{+}$ ($S_i^{-}$) is that of the dipole raising (lowering) operator. Besides, $\gamma_{ii}\equiv \gamma$ are the spontaneous emission rates of the atoms due to their interaction with the field, while $\gamma_{ij}$ and $\Omega_{ij}$ ($i\neq j$) describe the collective damping and dipole-dipole interaction potential, respectively. The first two terms on the right-hand side of Eq. \eqref{eq2-1} describe the unitary part of the evolution governed by the free Hamiltonian $H_i=\omega_0 S_i^z$ and the dipole-dipole interaction potential, and the last term results in the energy exchange between the system and the field. The explicit forms of $\gamma_{ij}$ and $\Omega_{ij}$ ($i\neq j$) are as follows \cite{ME0,ME1,ME2,ME3}:
\begin{equation}\label{eq2-2}
\begin{aligned}
 \gamma_{ij}= \,& \frac{3}{2}\gamma \Bigg\{ \big[1-(\hat{\mu}\cdot \hat{r}_{ij})^2\big] \frac{\sin(kr_{ij})}{kr_{ij}} \\
                & +\big[1-3(\hat{\mu}\cdot \hat{r}_{ij})^2\big]
                   \Bigg[\frac{\cos(kr_{ij})}{(kr_{ij})^2}-\frac{\sin(kr_{ij})}{(kr_{ij})^3}\Bigg]\Bigg\}, \\
 \Omega_{ij}= \,& \frac{3}{4}\gamma \Bigg\{-\big[1-(\hat{\mu}\cdot \hat{r}_{ij})^2\big] \frac{\cos(kr_{ij})}{kr_{ij}} \\
                & +\big[1-3(\hat{\mu}\cdot \hat{r}_{ij})^2\big]
                   \Bigg[\frac{\sin(kr_{ij})}{(kr_{ij})^2}+\frac{\cos(kr_{ij})}{(kr_{ij})^3}\Bigg]\Bigg\},
\end{aligned}
\end{equation}
where $k= \omega_0/c= 2\pi/\lambda$ is the wave vector ($c$ and $\lambda$ denote the velocity of light and the atomic resonant wavelength, respectively), $\hat{\mu}$ is the unit vector along the transition dipole moments of the two atoms that we assume are parallel to each other (i.e., $\hat{\mu}_1= \hat{\mu}_2 \equiv \hat{\mu}$), while $r_{ij}= |\vec{r}_{ij}|$ is the distance of the atoms and $\hat{r}_{ij}= \vec{r}_{ij}/r_{ij}$ is the unit vector along the interatomic axis.

For the charger-battery model of Eq. \eqref{eq2-1}, we are interested in analyzing the charging performance. To this aim, we consider the ergotropy, which quantifies the maximal amount of work that can be reversibly extracted from a QB. For a given QB with the free Hamiltonian $H$ and initial state $\rho$, the average work extractable from it via a cyclic unitary transformation $U$ is $\mathcal{W}(\rho,U)= \tr[(\rho -U\rho U^\dag) H]$, and the ergotropy $\mathcal{E}$ is defined as the maximum of $\mathcal{W}$ over the set of $U\in \mathcal{U}_c$ in the Hilbert space $\mathcal{H}_d$ ($d=\dim\rho$), i.e.,
\begin{equation}\label{eq2-3}
 \mathcal{E}(\rho)= \max_{U\in \mathcal{U}_c} \mathcal{W}(\rho,U),
\end{equation}
where $\mathcal{U}_c$ can be generated in a given time interval by applying suitable control fields to the system.

The optimal state $\tilde{\rho}= \tilde{U}\rho\tilde{U}^\dag$, realizing the maximum in Eq. \eqref{eq2-3}, is called the passive state associated with $\rho$. By rewriting $\rho$ in its eigenbasis as $\rho= \sum_j r_j |r_j\rangle \langle r_j|$ and $H$ in its eigenbasis as $H= \sum_k \varepsilon_k |\varepsilon_k\rangle \langle \varepsilon_k|$, where their respective eigenvalues are reordered as $r_j \geqslant r_{j+1}$ ($\forall j$) and $\varepsilon_k \leqslant \varepsilon_{k+1}$ ($\forall k$), the optimal unitary can be obtained as $\tilde{U}=\sum_j |\varepsilon_j\rangle \langle r_j|$, and the corresponding passive state is $\tilde{\rho}= \sum_j r_j |\varepsilon_j\rangle \langle \varepsilon_j|$. The ergotropy $\mathcal{E}$ can then be obtained explicitly as \cite{EPL}
\begin{equation}\label{eq2-4}
\mathcal{E}(\rho)=  \sum_k \varepsilon_k (\rho_{kk}-r_k),
\end{equation}
where $\rho_{kk}= \sum_j r_j |\langle r_j|\varepsilon_k\rangle|^2$ is the $k$th diagonal element of $\rho$ in the energy eigenbasis $\{|\varepsilon_k\rangle\}$ of $H$.

The ergotropy can be divided into two components. The incoherent component $\mathcal{E}_i(\rho)= \max_{V\in \mathcal{U}_c} \mathcal{W}(\rho,V)$ quantifies the maximum work extractable from a QB via coherence preserving operations $\mathcal{U}_c^{(i)}$, while the coherent component $\mathcal{E}_c(\rho)= \mathcal{E}(\rho)-\mathcal{E}_i(\rho)$ quantifies the work which cannot be extracted via only incoherent operations. After optimizing over all $V\in \mathcal{U}_c$, one can obtain \cite{resource6}
\begin{equation}\label{eq2-5}
\mathcal{E}_i(\rho)=  \sum_k \varepsilon_k (\rho_{kk}-\rho_{\tilde{\pi}_k \tilde{\pi}_k}),~
\mathcal{E}_c(\rho)= \sum_k \varepsilon_k  (\rho_{\tilde{\pi}_k \tilde{\pi}_k}-r_k),
\end{equation}
where $\{\rho_{\tilde{\pi}_k \tilde{\pi}_k}\}_{k=1,\dots, d}$ is the rearrangement of $\{\rho_{kk}\}_{k=1,\dots, d}$ in decreasing order. $\mathcal{E}_i(\rho)$ also equals to the ergotropy $\mathcal{E}(\delta_\rho)$ of the dephased state $\delta_\rho= \mathrm{diag} \{\rho_{11}, \ldots, \rho_{dd}\}$ \cite{shihl2}.

\section{Charging performance of the QB} \label{sec:3}
In this work, we consider the charging performance of a QB for two scenarios. The first scenario refers to the case in which the charger energy is not always maximal and the QB is empty initially, while the second one refers to the case in which the charger energy is maximal (i.e., it is in the fully excited state $|1\rangle$) and the QB is active (i.e., it has residual ergotropy) initially. The forms of the initial charger and battery states that we are going to discuss are as follows:
\begin{equation}\label{eq3-1}
 |\psi\rangle_{\mathrm{ch}}= \sqrt{c}|1\rangle + \sqrt{1-c} |0\rangle,~
 |\varphi\rangle_{\mathrm{ba}}= \sqrt{e}|1\rangle + \sqrt{1-e} |0\rangle,
\end{equation}
and we will compare the corresponding charging performance with that achieved via the diagonal forms of the initial charger and battery states given by
\begin{equation}\label{eq3-2}
 \rho_{\mathrm{diag,ch}}= \mathrm{diag}\{c,1-c\},~
 \rho_{\mathrm{diag,ba}}= \mathrm{diag}\{e,1-e\},
\end{equation}
where $c\in[0,1]$ and $e\in[0,1]$ in the above two equations.

For a general initial state of the two atoms, e.g., $|\psi\rangle_{\mathrm{ch}}\otimes |\varphi\rangle_{\mathrm{ba}}$, one can solve numerically the master equation \eqref{eq2-1}, while for the initial X state, that is, the density matrix $\rho(0)$ with nonzero elements only along the main diagonal and anti-diagonal (e.g., $\rho_{\mathrm{diag,ch}}\otimes \rho_{\mathrm{diag,ba}}$), analytical solution of the master equation \eqref{eq2-1} can also be obtained \cite{xstate}. Having $\rho(t)$ at hand, we can obtain $\mathcal{E}_{\mathrm{ba}}$ of the QB and other relevant quantities (see below) characterizing the charging performance. We will show that the spontaneous emission, collective damping, and dipole-dipole interaction of the atoms, as expected, are key in determining the charging performance, which can be significantly enhanced by a proper choice of their strengths.

For the charger-battery model in Sec. \ref{sec:2}, the charging process can be implemented as follows. Initially, the interaction of the two elements (i.e., the charger and the QB) with the vacuum field is switched off, and the ergotropies for the battery states of Eqs. \eqref{eq3-1} and \eqref{eq3-2} are given by $\mathcal{E}_{\mathrm{ba}}(|\varphi\rangle_{\mathrm{ba}})=e\omega_0$ and $\mathcal{E}_{\mathrm{ba}}(\rho_{\mathrm{diag,ba}}) =\max\{0,(2e-1)\omega_0\}$. When $t>0$, the interaction between the two elements and the multimode vacuum field is switch on. In this way, the charging process begins and the evolution of the system can be engineered by choosing proper control parameters in order to lead the QB towards a state that is most active, that is, there is as much extractable work as possible in it. Here, we define the dynamical maximum of $\mathcal{E}_{\mathrm{ba}}$ (which we denote it by $\bar{\mathcal{E}}_{\mathrm{ba}}$) as the ergotropy charged on the QB, and the interaction time $\bar{t}$ it takes to charge the QB up to its dynamical maximum as the charging time. To avoid further energy dissipation to the field and energy backflow to the charger, the interaction of the QB to the field is switched off after $t>\bar{t}$, that is, the charging process ends when the QB reaches the intermediate state $\rho_{\mathrm{ba}}(\bar{t})$. Notably, this process is somewhat different from that of Refs. \cite{collective5,noiseqb1}, as in that case the steady state, which is of equilibrium and independent of the initial conditions, corresponds to a charged battery. But the charged battery $\rho_{\mathrm{ba}}(\bar{t})$ at the intermediate time $\bar{t}$ strongly depends on the initial states of the system.

\begin{figure}
\centering
\resizebox{0.45 \textwidth}{!}{%
\includegraphics{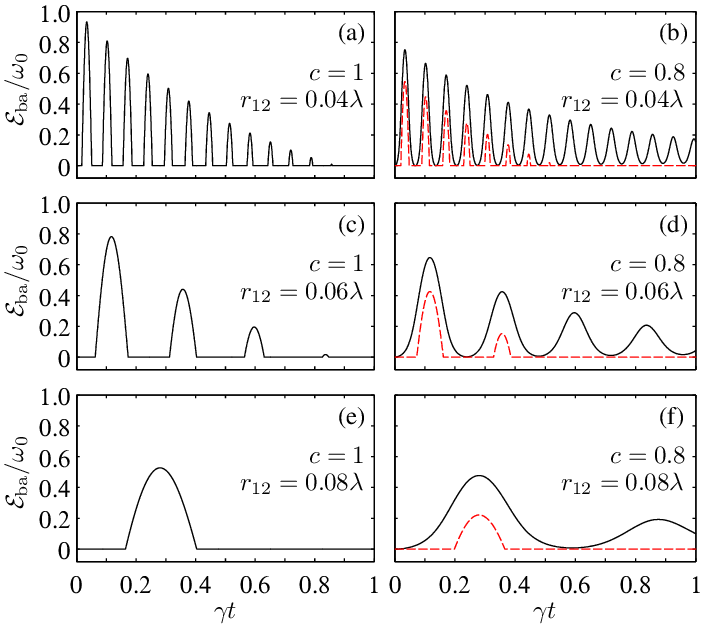}}
\caption{Dynamics of $\mathcal{E}_{\mathrm{ba}}$ of the QB with $\hat{\mu}\cdot \hat{r}_{12}=0$ and different $c$ and $r_{12}$, where the charger is in the initial state $|\psi\rangle_{\mathrm{ch}}$ (solid black) or $\rho_{\mathrm{diag,ch}}$ (dashed red), and the QB is in the initial state $|0\rangle$.} \label{fig:1}
\end{figure}

To begin with, we consider the scenario in which the QB is in its ground state initially. In Fig. \ref{fig:1} we showcase the evolution of $\mathcal{E}_{\mathrm{ba}}$ of the QB charged by a charger in the initial state $|\psi\rangle_{\mathrm{ch}}$ or $\rho_{\mathrm{diag,ch}}$, where we have assumed that the atomic dipole moments are polarized in the direction perpendicular to the interatomic axis, i.e., $\hat{\mu}\cdot \hat{r}_{12}=0$. When  $|\psi\rangle_{\mathrm{ch}}=|1\rangle$ (thereby its energy is maximal), one can obtain
\begin{equation}\label{eq3-3}
 \mathcal{E}_{\mathrm{ba}}=\max\{0,e^{-\gamma t}[\cosh(\gamma_{12}t)-\cos(2\Omega_{12}t)]-1\},
\end{equation}
hence the evolution of the ergotropy $\mathcal{E}_{\mathrm{ba}}$ is profoundly affected by the collective damping and the dipole-dipole interaction. As anticipated in Eq. \eqref{eq3-3} and shown in the left three panels of Fig. \ref{fig:1}, $\mathcal{E}_{\mathrm{ba}}$ first remains 0 for a period of time $t_0$, after which it progressively increases and reaches its maximum $\bar{\mathcal{E}}_{\mathrm{ba}}$ at $t=\bar{t}$. Both $t_0$ and $\bar{t}$ can be obtained numerically from Eq. \eqref{eq3-3}. The collective damping induces decay of the peak values of  $\mathcal{E}_{\mathrm{ba}}$, whereas the dipole-dipole interaction manifests its presence in the oscillatory behaviors of $\mathcal{E}_{\mathrm{ba}}$. One can also note that $\bar{\mathcal{E}}_{\mathrm{ba}}$ decreases with the increasing distance $r_{12}$ between the charger and the QB, whereas $t_0$ and $\bar{t}$ increase with the increasing $r_{12}$. Moreover, there are multiple time intervals (dark periods) at which $\mathcal{E}_{\mathrm{ba}}$ vanishes, which can also be explained from Eq. \eqref{eq3-3}. When the charger is in a superposition of the excited and ground states (thereby its energy is not maximal initially), the ergotropy $\mathcal{E}_{\mathrm{ba}}$ increases with the evolving time $t$ from the very beginning. However, this is not the case for $\rho_{\mathrm{diag,ch}}$ with the same $c$ (which has the same energy as $|\psi\rangle_{\mathrm{ch}}$), as in this case $\mathcal{E}_{\mathrm{ba}}$ remains 0 for a period of time, and the ergotropy charged on the QB is smaller than that charged with the initial charger state $|\psi\rangle_{\mathrm{ch}}$. A further calculation reveals that $\mathcal{E}_{\mathrm{ba}}$ for the initial charger state $\rho_{\mathrm{diag,ch}}$ equals to the incoherent component of $\mathcal{E}_{\mathrm{ba}}$ for the initial charger state $|\psi\rangle_{\mathrm{ch}}$. This indicates that the discrepancy between the solid and dashed lines in the right three panels of Fig. \ref{fig:1} comes from the coherent contribution to ergotropy. That is, the evolution of the system from $|\psi\rangle_{\mathrm{ch}}$ generates coherence in the energy eigenbasis, thereby enhances the ergotropy $\mathcal{E}_{\mathrm{ba}}$. Such an enhancement due to generation of coherence has also been reported in Refs. \cite{collective5,resource6}.

\begin{figure}
\centering
\resizebox{0.45 \textwidth}{!}{%
\includegraphics{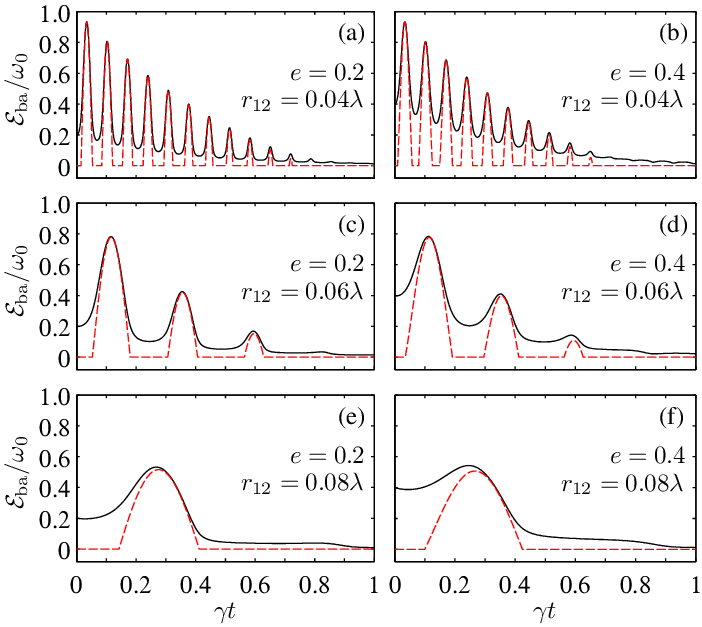}}
\caption{Dynamics of $\mathcal{E}_{\mathrm{ba}}$ of the QB with $\hat{\mu}\cdot \hat{r}_{12}=0$ and different $e$ and $r_{12}$, where the charger is in the initial state $|1\rangle$, and the QB is in the initial state $|\varphi\rangle_{\mathrm{ba}}$ (solid black) or $\rho_{\mathrm{diag,ba}}$ (dashed red).} \label{fig:2}
\end{figure}

 When considering the scenario in which the charger energy is maximal (i.e., it is in the excited state) and the QB is active initially, we showcase in Fig. \ref{fig:2} the evolution of $\mathcal{E}_{\mathrm{ba}}$ of the QB in the initial state $|\varphi\rangle_{\mathrm{ba}}$ by the solid black lines, where we have also taken $\hat{\mu}\cdot \hat{r}_{12}=0$. One can see that $\mathcal{E}_{\mathrm{ba}}$ increases with time from the very beginning and reaches its dynamical maximum $\bar{\mathcal{E}}_{\mathrm{ba}}$ at $t=\bar{t}$. Here, $\bar{\mathcal{E}}_{\mathrm{ba}}$ ($\bar{t}$) also decreases (increases) with the increase of the distance $r_{12}$ between the charger and QB. For the fixed $r_{12}$, a detailed comparison shows that $\bar{\mathcal{E}}_{\mathrm{ba}}$ ($\bar{t}$) can only be slightly enhanced (shortened) by increasing $e$ properly. But it should be note that when $e$ is larger than a critical value, the QB will cannot be charged as $\bar{\mathcal{E}}_{\mathrm{ba}}$ becomes smaller than the initial ergotropy $e\omega_0$ in it (we will discuss this issue in detail later). For the case that the QB is in the initial diagonal state $\rho_{\mathrm{diag,ba}}$ with $e<0.5$, as shown by the dashed red lines in Fig. \ref{fig:2}, $\mathcal{E}_{\mathrm{ba}}$ remains zero for a time interval, after which it increases to its dynamical maximum $\bar{\mathcal{E}}_{\mathrm{ba}}$, which is only slightly less than that achieved with the initial battery state $|\varphi\rangle_{\mathrm{ba}}$.

\begin{figure}
\centering
\resizebox{0.45 \textwidth}{!}{%
\includegraphics{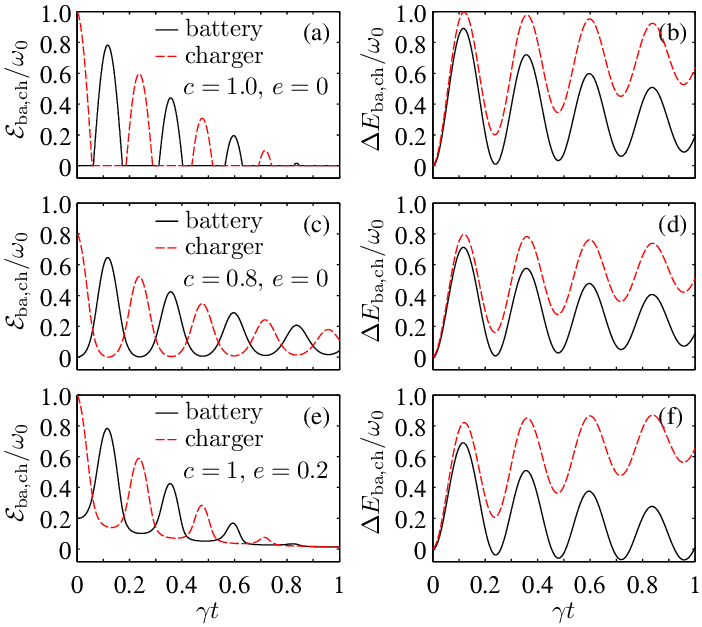}}
\caption{(Left) Comparison of $\mathcal{E}_{\mathrm{ba}}$ of the QB (solid black) and $\mathcal{E}_{\mathrm{ch}}$ of the charger (dashed red) with $\hat{\mu}\cdot \hat{r}_{12} =0$, $r_{12}=0.06\lambda$, and different $|\psi\rangle_{\mathrm{ch}}$ and $|\varphi\rangle_{\mathrm{ba}}$. (Right) Comparison of the input energy $\Delta E_{\mathrm{ba}}$ of the QB (solid black) and output energy $\Delta E_{\mathrm{ch}}$ of the charger (dashed red) with the same parameters as those in the left three panels.} \label{fig:3}
\end{figure}

To reveal more details about the charging process, we further provide in Fig. \ref{fig:3} a comparison of $\mathcal{E}_{\mathrm{ba}}$ of the QB and $\mathcal{E}_{\mathrm{ch}}$ of the charger, as well as the input energy $\Delta E_{\mathrm{ba}}=E_{\mathrm{ba}}(t)-E_{\mathrm{ba}}(0)$ of the QB and the output energy $\Delta E_{\mathrm{ch}}=E_{\mathrm{ch}}(0)-E_{\mathrm{ch}}(t)$ of the charger. When the charger and QB are in the initial states $|1\rangle$ and  $|0\rangle$, respectively, we find that
\begin{equation}\label{eq3-4}
\begin{aligned}
 & \mathcal{E}_{\mathrm{ch}}=\max\{0,e^{-\gamma t}[\cosh(\gamma_{12}t)+\cos(2\Omega_{12}t)]-1\}, \\
 & \Delta E_{\mathrm{ba}}= \frac{1}{2}e^{-\gamma t}[\cosh(\gamma_{12}t)-\cos(2\Omega_{12}t)], \\
 & \Delta E_{\mathrm{ch}}= 1-\frac{1}{2}e^{-\gamma t}[\cosh(\gamma_{12}t)+\cos(2\Omega_{12}t)],
\end{aligned}
\end{equation}
and as shown in Fig. \ref{fig:3}(a), the ergotropy $\mathcal{E}_{\mathrm{ba}}$ of the QB does not increase synchronously with the decrease of the ergotropy $\mathcal{E}_{\mathrm{ch}}$ of the charger. Specifically, $\mathcal{E}_{\mathrm{ch}}$ decreases from $\omega_0$ to its minimum 0 at $t=t_0$, and it is from the same time $t_0$ that the ergotropy $\mathcal{E}_{\mathrm{ba}}$ increases from 0 to its first dynamical maximum $\bar{\mathcal{E}}_{\mathrm{ba}}$. For the other two cases, as shown in Fig. \ref{fig:3}(c) and (e), $\mathcal{E}_{\mathrm{ba}}$ of the QB increases from the very beginning and reaches its first dynamical maximum at $t=\bar{t}$, which is accompanied by a synchronously decrease of $\mathcal{E}_{\mathrm{ch}}$. Of course, a detailed comparison shows that $\mathcal{E}_{\mathrm{ch}}$ continuously decreases for a very short period of time after $t=\bar{t}$. The different behaviors of $\mathcal{E}_{\mathrm{ba}}$ are due to the different contributions of the two components (i.e., the incoherent component $\mathcal{E}_{\mathrm{ba},i}$ and coherent component $\mathcal{E}_{\mathrm{ba},c}$) of the ergotropy. In fact, for the case in Fig. \ref{fig:3}(a), $\mathcal{E}_{\mathrm{ba},c}$ remains 0 in the whole time region, whereas for that in Fig. \ref{fig:3}(c) and (e), $\mathcal{E}_{\mathrm{ba},i}$ remains 0 for a short time interval before increasing to its maximum and $\mathcal{E}_{\mathrm{ba},c}$ increases from the very beginning (for the conciseness of this paper, we do not show the plots here). As for the energy exchange, from the right three panels of Fig. \ref{fig:3} one can see that the energy is injected into the QB from the beginning, most of which could be converted into extractable work. Of course, it is inevitable that partial of the energy output from the charger is dissipated as heat, thereby one always has $\Delta E_{\mathrm{ba}}<\Delta E_{\mathrm{ch}}$. Moreover, as $[S_1^z+S_2^z,S_i^{+}S_j^{-}]=0$ [see Eq. \eqref{eq2-1}], the dissipated heat $Q$ is independent of the dipole-dipole interaction, and for the special case shown in Fig. \ref{fig:3}(b), using Eq. \eqref{eq3-4} we obtain $Q= [1-e^{-\gamma t} \cosh(\gamma_{12}t)]\omega_0$, which grows monotonically in time.

We have also considered the initial charger state $\rho_{\mathrm{diag,ch}}$ of Eq. \eqref{eq3-2} with $c\leqslant 0.5$. For this case, the initial charger energy is $(c-0.5)\omega_0$ and its ergotropy is 0. But our calculation shows that for any initial state of the QB (including but not limited to $|\varphi\rangle_{\mathrm{ba}}$ and $\rho_{\mathrm{diag,ba}}$), while there is considerable amount of energy being transferred to the QB (i.e., $\Delta E_{\mathrm{ba}}>0$), the ergotropy gain $\mathcal{E}_{\mathrm{ba}}(t)-\mathcal{E}_{\mathrm{ba}}(0)$ of it remains 0, irrespective of $r_{12}$ and $\hat{\mu}\cdot \hat{r}_{12}$. In this sense, it seems that it is the charger ergotropy instead of its energy that determines the ergotropy charged on the QB. A physical reason for this is that the collective extractable work for the charger-battery system cannot be increased as there is no external driving in the system Hamiltonian.

\begin{figure}
\centering
\resizebox{0.45 \textwidth}{!}{%
\includegraphics{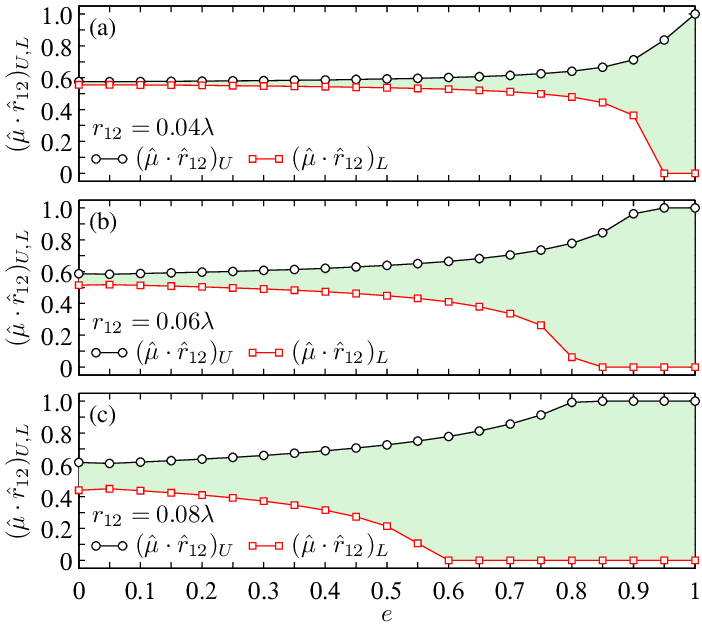}}
\caption{The $e$ dependence of the bounds $(\hat{\mu}\cdot \hat{r}_{12})_U$ and $(\hat{\mu}\cdot \hat{r}_{12})_L$ with different $r_{12}$, where the charger and QB are in the initial states $|1\rangle$ and $|\varphi\rangle_{\mathrm{ba}}$, respectively. The QB could be charged only when $\hat{\mu}\cdot \hat{r}_{12}$ locates outside the green shaded region.} \label{fig:4}
\end{figure}

So far we have examined the case of $\hat{\mu}\cdot \hat{r}_{12}=0$, that is, the atomic dipole moments are polarized in the direction perpendicular to the interatomic axis. But if $\hat{\mu}\cdot \hat{r}_{12}\neq 0$, the QB may cannot always be charged. For example, for the initial states in Eq. \eqref{eq3-1}, there exists a region $\hat{\mu}\cdot \hat{r}_{12} \in [(\hat{\mu}\cdot \hat{r}_{12})_L, (\hat{\mu}\cdot \hat{r}_{12})_U]$ in which the QB cannot be charged under certain initial conditions. By choosing $|\psi\rangle_{\mathrm{ch}}= |1\rangle$, we solved numerically the  bounds $(\hat{\mu}\cdot \hat{r}_{12})_L$ and $(\hat{\mu}\cdot \hat{r}_{12})_U$ for three different distances $r_{12}$ between the charger and the QB. As can be seen from Fig. \ref{fig:4}, such a region expands with the increase of $e$. When $e$ is comparable to its maximum 1, the QB cannot be charged in nearly the whole region of $\hat{\mu}\cdot \hat{r}_{12}$. Moreover, the region also expands with an increase in $r_{12}$; in particular, the lower bound $(\hat{\mu}\cdot \hat{r}_{12})_L$ approaches to its minimum 0 faster than the rate for which the upper bound $(\hat{\mu}\cdot \hat{r}_{12})_U$ approaches to its maximum 1. When $r_{12}\gtrsim 0.1691\lambda$, the region becomes $\hat{\mu}\cdot \hat{r}_{12} \in [0,1]$ and the QB cannot be charged any more. So in order to achieve an optimal charging performance, one should choose carefully the magnitudes of $\hat{\mu} \cdot \hat{r}_{12}$ and $r_{12}$. It is also better to charge the QB after it is fully discharged.

\begin{figure}
\centering
\resizebox{0.45 \textwidth}{!}{%
\includegraphics{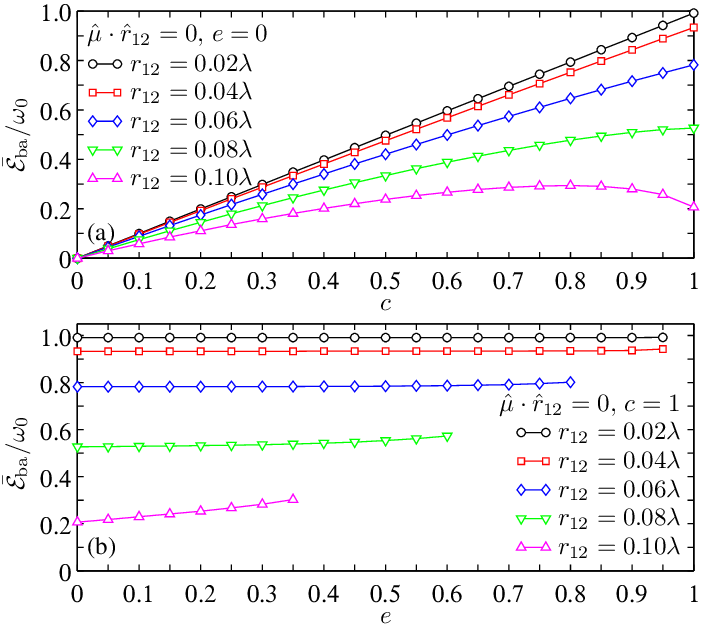}}
\caption{The ergotropy $\bar{\mathcal{E}}_{\mathrm{ba}}$ charged on the QB at time $\bar{t}$ with $\hat{\mu}\cdot \hat{r}_{12}=0$ and different $r_{12}$. (a) [(b)] The charger and QB are in the initial states $|\psi\rangle_{\mathrm{ch}}$ and $|0\rangle$ ($|1\rangle$ and $|\varphi\rangle_{\mathrm{ba}}$), respectively. The lines in (b) are plotted in different regions of $e$ as the QB cannot be charged when $e$ is larger than a $r_{12}$-dependent critical value.} \label{fig:5}
\end{figure}

To gain more insight into the performance of the charger-battery model, we further analyze how the initial charger energy and the residual ergotropy in the QB affect the maximal ergotropy $\bar{\mathcal{E}}_{\mathrm{ba}}$. As shown in Fig. \ref{fig:5}(a), when the charger and QB are in the initial states $|\psi\rangle_{\mathrm{ch}}$ and $|0\rangle$, respectively, $\bar{\mathcal{E}}_{\mathrm{ba}}$ increases monotonically from 0 to its maximum; in particular, for small enough $r_{12}$ (e.g., $r_{12}=0.02\lambda$), it increases nearly linearly with the increase of $c$. For the moderate values of $r_{12}$ (e.g., $r_{12}=0.10\lambda$), $\bar{\mathcal{E}}_{\mathrm{ba}}$ first increases to a maximum and then decreases to its minimum. Of course, as we have pointed out previously, for $r_{12}\gtrsim 0.1691\lambda$, $\bar{\mathcal{E}}_{\mathrm{ba}}$ will remain 0 in the whole region of $e$. When the charger and QB are in the initial states $|1\rangle$ and $|\varphi\rangle_{\mathrm{ba}}$, respectively, as shown in Fig. \ref{fig:5}(b), the QB cannot be charged (here, by saying a QB cannot be charged, we mean that $\bar{\mathcal{E}}_{\mathrm{ba}}$ is smaller than the initial ergotropy in it) if $e$ is larger than a $r_{12}$-dependent threshold, while out of this region, $\bar{\mathcal{E}}_{\mathrm{ba}}$ is weakly dependent on $e$ when $r_{12}$ is very small. That is, in this case, there is a trade-off between the initial ergotropy $e\omega_0$ in the QB and its ergotropy gain $\bar{\mathcal{E}}_{\mathrm{ba}}- e\omega_0$, meaning that the latter decreases with the increase of $e$. In fact, when $r_{12}$ is very small, it can be obtained from Eq. \eqref{eq2-2} that $|\Omega| \gg |\gamma_{12}|$. Hence in the short time region the dissipative term in Eq. \eqref{eq2-1} can be neglected, and the solution of this master equation can be approximated as $\rho(t) \simeq U\rho(0) U^\dagger$, where
\begin{equation}\label{eq3-5}
U=
  \left(\begin{array}{cccc}
    e^{-i\omega_0 t} & 0 & 0 & 0; \\
    0  & \cos(\Omega_{12}t)  & -i\sin(\Omega_{12}t)   & 0 \\
    0  & -i\sin(\Omega_{12}t)   & \cos(\Omega_{12}t)  & 0 \\
    u  & 0   & 0   & e^{i\omega_0 t} \\
  \end{array}\right),
\end{equation}
then for the initial charger state $|1\rangle$ and battery state $|\varphi\rangle_{\mathrm{ba}}$ of Eq. \eqref{eq3-1}, that is, the case in Fig. \ref{fig:5}(b), it can be shown that
\begin{equation}\label{eq3-7}
\mathcal{E}_{\mathrm{ba,ap}}=\frac{1}{2} \max\left\{0,\left[\alpha+\sqrt{\alpha^2+4 e(1-e)\cos^2(\Omega_{12}t)}\right]\omega_0\right\},
\end{equation}
where $\alpha=e-(1-e)\cos(2\Omega_{12}t)$, and we denote the ergotropy as $\mathcal{E}_{\mathrm{ba,ap}}$ for distinguishing it from $\mathcal{E}_{\mathrm{ba}}$ achieved without approximation. It can be seen that $\mathcal{E}_{\mathrm{ba,ap}}$ takes its maximum $\omega_0$ at $t_{\mathrm{ap}} =\pi/(2|\Omega_{12}|)$. Of course, this is an approximation and the error grows with the increasing $r_{12}$; for example, for $\hat{\mu}\cdot \hat{r}_{12}=0$ and $e=0.2$, $|\bar{t}-t_{\mathrm{ap}}|/\bar{t} \simeq 1.72 \times 10^{-4}$ and $|\bar{\mathcal{E}}_{\mathrm{ba}}-\omega_0|/\bar{\mathcal{E}}_{\mathrm{ba}} \simeq 0.0010$ when $r_{12}=0.01\lambda$, while they turn out to be $0.0112$ and $0.0713$ when $r_{12}=0.04\lambda$.

\begin{figure}
\centering
\resizebox{0.45 \textwidth}{!}{%
\includegraphics{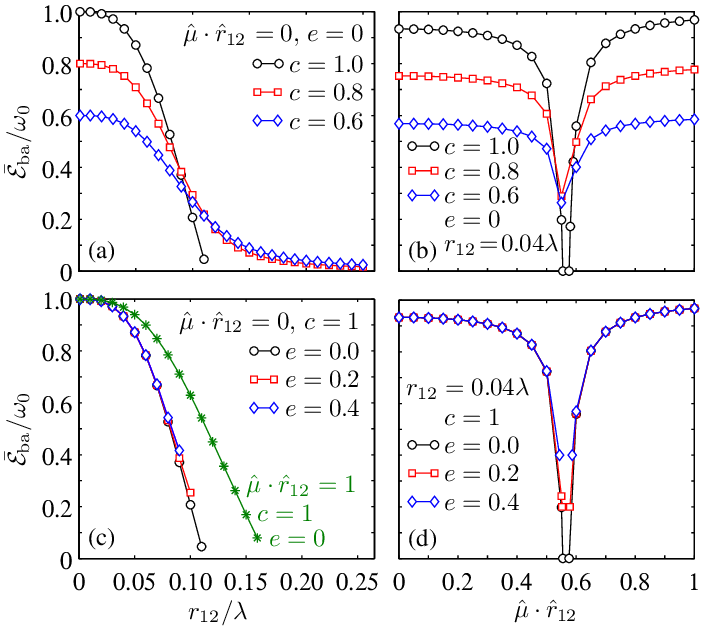}}
\caption{The $r_{12}$ ($\hat{\mu}\cdot \hat{r}_{12}$) dependence of $\bar{\mathcal{E}}_{\mathrm{ba}}$ charged on the QB at time $\bar{t}$ with $\hat{\mu}\cdot \hat{r}_{12}=0$ ($r_{12}=0.04\lambda$) and different $c$ and $e$, where the charger and QB are in the initial states $|\psi\rangle_{\mathrm{ch}}$ and $|\varphi\rangle_{\mathrm{ba}}$, respectively. The lines are plotted in the parameter regions in which the QB can be charged, and the first point in (a) and (c) correspond to $r_{12}=0.001\lambda$. The green asterisks in (c) are plotted with $\hat{\mu}\cdot \hat{r}_{12}=1$, $c=1$, and $e=0$.} \label{fig:6}
\end{figure}

In Fig. \ref{fig:6}(a) and (c), we showcase the $r_{12}$ dependence of $\bar{\mathcal{E}}_{\mathrm{ba}}$ with $\hat{\mu}\cdot \hat{r}_{12}=0$. There we can see that when the distance $r_{12}$ is very short, most of the charger energy could be transferred to the QB and converted into extractable work. When the initial charger energy is maximal, the QB is almost fully charged at $r_{12} \rightarrow 0$, and $\bar{\mathcal{E}}_{\mathrm{ba}}$ decreases with the increase of $r_{12}$ and vanishes after $r_{12}$ exceeds a threshold. If the initial charger energy is not maximal, $\bar{\mathcal{E}}_{\mathrm{ba}}$ also decreases with the increase of $r_{12}$. But in this case, it can take a finite value in an extended region of $r_{12}$, and such a small difference is also due to the generation of coherence in the system dynamics. In Fig. \ref{fig:6}(b) and (d) we show the $\hat{\mu}\cdot \hat{r}_{12}$ dependence of $\bar{\mathcal{E}}_{\mathrm{ba}}$ with $r_{12}=0.04\lambda$, where the lines are discontinuous as there are regions of $\hat{\mu}\cdot \hat{r}_{12}$ in which the QB cannot be charged (see Fig. \ref{fig:4}). It is evident that a very small or large $\hat{\mu}\cdot \hat{r}_{12}$ is beneficial for improving the charging performance. Moreover, similar to that displayed in Fig. \ref{fig:5}(b), one can note from the bottom two panels of Fig. \ref{fig:6} that $\bar{\mathcal{E}}_{\mathrm{ba}}$ is weakly dependent on the initial ergotropy $e\omega_0$ in the QB. This confirms again that the ergotropy gain of the QB decreases with the increase of $e\omega_0$ in a wide region of $r_{12}$ and $\hat{\mu}\cdot \hat{r}_{12}$, although the initial charger energy is always maximal. So compared to the process starting from a passive state (i.e., $e=0$), the process starting from an active state produces more dissipation and thereby has a lower efficiency. From Figs. \ref{fig:6}(b) and (d) one can also see that $\hat{\mu}\cdot \hat{r}_{12}\sim 1$ is slightly more efficient than that of $\hat{\mu} \cdot \hat{r}_{12}\sim 0$ in enhancing $\bar{\mathcal{E}}_{\mathrm{ba}}$. To fully understand this effect, we further plotted in Figs. \ref{fig:6}(c) the $r_{12}$ dependence of $\bar{\mathcal{E}}_{\mathrm{ba}}$ with $\hat{\mu}\cdot \hat{r}_{12}= 1$ (denoted by the green asterisk). As seen, for small $r_{12}$, $\bar{\mathcal{E}}_{\mathrm{ba}}$ for $\hat{\mu}\cdot \hat{r}_{12}= 0$ and 1 are approximately the same. But with the increasing $r_{12}$, the difference between them becomes more and more apparent, and the region of $r_{12}$ in which $\bar{\mathcal{E}}_{\mathrm{ba}}$ takes a finite value is also extended for $\hat{\mu}\cdot \hat{r}_{12}= 1$.

\begin{figure}
\centering
\resizebox{0.45 \textwidth}{!}{%
\includegraphics{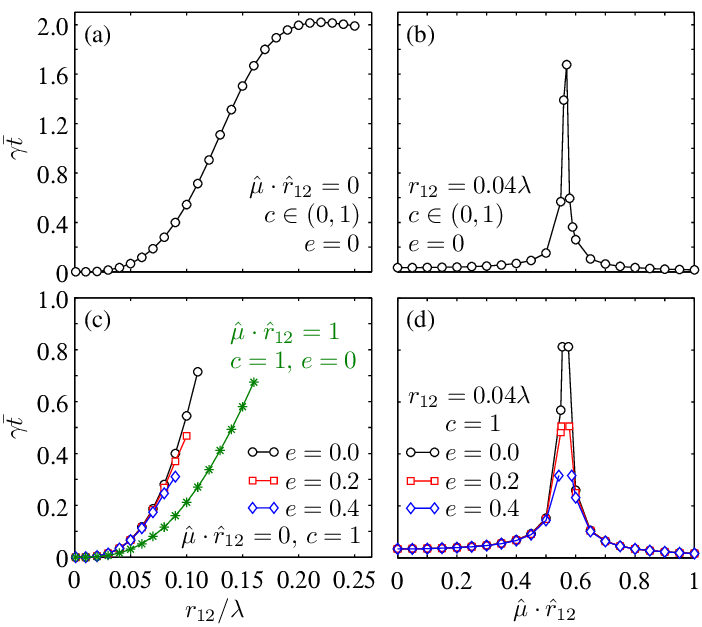}}
\caption{The $r_{12}$ ($\hat{\mu}\cdot \hat{r}_{12}$) dependence of the charging time $\bar{t}$ with $\hat{\mu}\cdot \hat{r}_{12}=0$ ($r_{12}=0.04\lambda$) and different $c$ and $e$, where the charger and QB are in the initial states $|\psi\rangle_{\mathrm{ch}}$ and $|\varphi\rangle_{\mathrm{ba}}$, respectively. The lines are plotted in the parameter regions in which the QB can be charged, and the first point in (a) and (c) correspond to $r_{12}=0.001\lambda$. The green asterisks in (c) are plotted with $\hat{\mu}\cdot \hat{r}_{12}=1$, $c=1$, and $e=0$.} \label{fig:7}
\end{figure}

Apart from $\bar{\mathcal{E}}_{\mathrm{ba}}$, another figure of merit related to the performance of a QB is the charging time $\bar{t}$. As we said previously, $\bar{t}$ is the time it takes to charge the QB with $\mathcal{E}_{\mathrm{ba}}(\bar{t})= \bar{\mathcal{E}}_{\mathrm{ba}}$. For the charger-battery model given in Sec. \ref{sec:2}, we showcase in Fig. \ref{fig:7} how $r_{12}$ and $\hat{\mu}\cdot \hat{r}_{12}$ affect $\gamma\bar{t}$ under different initial conditions. For the case of fixed $\hat{\mu}\cdot \hat{r}_{12}=0$, one can note that a short distance $r_{12}$ between the charger and QB helps to realizing the fast charging. The charging time $\gamma\bar{t}$ is independent of the initial charger energy if the QB is in its ground state $|0\rangle$ initially, while it is weakly dependent on the initial ergotropy $e\omega_0$ of the QB if the charger energy is maximal initially. This is different from that of a classical battery, for which the charging time could be shortened when there is residual energy in it. Moreover, one can see from Fig. \ref{fig:7}(b) and (d) (the discontinuities of the lines are also due to the fact that there are regions for $\hat{\mu}\cdot \hat{r}_{12}$ in which the QB cannot be charged) that a small or large $\hat{\mu}\cdot \hat{r}_{12}$ is beneficial for shortening the charging time, while in the boundary regions of $\hat{\mu}\cdot \hat{r}_{12}\in [(\hat{\mu}\cdot \hat{r}_{12})_L, (\hat{\mu}\cdot \hat{r}_{12})_U]$, the charging time becomes very long, thus it is disadvantageous for improving the charging performance. Besides, in Fig. \ref{fig:7}(c) we further plotted the $r_{12}$ dependence of $\gamma\bar{t}$ with $\hat{\mu}\cdot \hat{r}_{12}= 1$ (see the green asterisk). Similar to that of the ergotropy shown in Fig. \ref{fig:6}(c), the difference of $\gamma\bar{t}$ for $\hat{\mu}\cdot \hat{r}_{12}= 0$ and 1 also increases with the increase of $r_{12}$, indicating that the large $\hat{\mu}\cdot \hat{r}_{12}$ is more efficient than that of a small one in improving the charging performance, especially when the charger-battery distance $r_{12}$ is relatively far.

\begin{figure}
\centering
\resizebox{0.45 \textwidth}{!}{%
\includegraphics{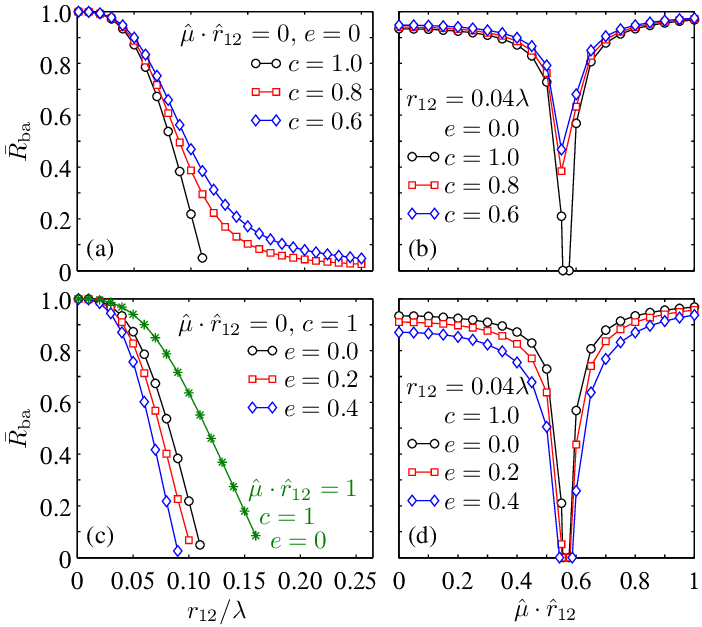}}
\caption{The $r_{12}$ ($\hat{\mu}\cdot \hat{r}_{12}$) dependence of $\bar{R}_{\mathrm{ba}}$ with $\hat{\mu}\cdot \hat{r}_{12}=0$ ($r_{12}=0.04\lambda$) and different $c$ and $e$, where the charger and QB are in the initial states $|\psi\rangle_{\mathrm{ch}}$ and $|\varphi\rangle_{\mathrm{ba}}$, respectively. The lines are plotted in the parameter regions in which the QB can be charged, and the first point in (a) and (c) correspond to $r_{12}=0.001\lambda$. The green asterisks in (c) are plotted with $\hat{\mu}\cdot \hat{r}_{12}=1$, $c=1$, and $e=0$.} \label{fig:8}
\end{figure}

In Fig. \ref{fig:8}, we further show the the charging efficiency $\bar{R}_{\mathrm{ba}}$ of the QB under different initial conditions. Here, $\bar{R}_{\mathrm{ba}}$ is defined as the proportion of output energy from the charger that could be converted into extractable work in the QB, i.e.,
\begin{equation}\label{eq3-7}
 \bar{R}_{\mathrm{ba}}= \frac{\Delta\bar{\mathcal{E}}_{\mathrm{ba}}}{\Delta \bar{E}_{\mathrm{ch}}}
                      = \frac{\max\{0,\bar{\mathcal{E}}_{\mathrm{ba}}(\bar{t})-\bar{\mathcal{E}}_{\mathrm{ba}}(0)\}}
                        {\tr\{[\rho_{\mathrm{ch}}(0)-\rho_{\mathrm{ch}}(\bar{t})]H_{\mathrm{ch}}\}},
\end{equation}
where $\Delta\bar{\mathcal{E}}_{\mathrm{ba}}$ is the net ergotropy charged on the QB, $\Delta \bar{E}_{\mathrm{ch}}$ is the output energy of the charger, and $H_{\mathrm{ch}}$ is the self-Hamiltonian of the charger. As can be seen from Fig. \ref{fig:8}(a) and (c), for the $\hat{\mu}\cdot \hat{r}_{12}=0$ case, $\bar{R}_{\mathrm{ba}}$ takes the value of about 1 when $r_{12}$ is very small and decreases with the increase of $r_{12}$, the phenomenon of which is very similar to that of $\bar{\mathcal{E}}_{\mathrm{ba}}$ showed in Fig. \ref{fig:6}. However, the dependence of $\bar{R}_{\mathrm{ba}}$ on $c$ and $e$ is opposite to that of $\bar{\mathcal{E}}_{\mathrm{ba}}$. Specifically, if the distance from the QB to the charger is not far, $\bar{\mathcal{E}}_{\mathrm{ba}}$ decreases with the increase of $c$ for $|\varphi\rangle_{\mathrm{ba}}= |0\rangle$ and is weakly dependent on $e$ for $|\psi\rangle_{\mathrm{ch}}=|1\rangle$, whereas the corresponding $\bar{R}_{\mathrm{ba}}$ is weakly dependent on $c$ for $|\varphi\rangle_{\mathrm{ba} }= |0\rangle$ and decreases with the increase of $e$ for $|\psi\rangle_{\mathrm{ch}}=|1\rangle$ (cf. the left two panels of Figs. \ref{fig:6} and \ref{fig:8}). Looking at Fig. \ref{fig:8}(b) and (d), one can note that a small or large $\hat{\mu}\cdot \hat{r}_{12}$ is also beneficial for enhancing the charging efficiency $\bar{R}_{\mathrm{ba}}$, and the dependence of $\bar{R}_{\mathrm{ba}}$ on $c$ and $e$ is also opposite to that of $\bar{\mathcal{E}}_{\mathrm{ba}}$ showed in the right two panels of Fig. \ref{fig:6}. Moreover, similar to $\bar{\mathcal{E}}_{\mathrm{ba}}$ and $\gamma\bar{t}$, the advantage of $\hat{\mu}\cdot \hat{r}_{12}=1$ in enhancing $\bar{R}_{\mathrm{ba}}$ becomes apparent in the relatively large $r_{12}$ region, as shown by the green asterisks in Fig. \ref{fig:8}(c).

Finally, it is worth remarking that the mean charging power $\bar{P}_{\mathrm{ba}}= \Delta\bar{\mathcal{E}}_{\mathrm{ba}}/\bar{t}$ can also be significantly enhanced by tuning the parameters in such a way that $r_{12}$ is very short and $\hat{\mu}\cdot \hat{r}_{12}=0$ or 1. The reason for this enhancement, as can be seen from Figs. \ref{fig:6} and \ref{fig:7}, is that in this case $\Delta\bar{\mathcal{E}}_{\mathrm{ba}}$ is significantly enhanced and $\bar{t}$ is greatly shortened. Moreover, for the chosen optimal $\hat{\mu}\cdot \hat{r}_{12}$ and $r_{12}$, $\bar{P}_{\mathrm{ba}}$ increases with the increase of the initial ergotropy in the charger; however, it decreases with the increase of the initial ergotropy in the QB.

\section{Summary} \label{sec:4}
In this paper, we have investigated a charger-battery model that consists of two atoms, where the first one is treated as the charger and the second one as the QB. Different figures of merit, such as the ergotropy, the charging time, and the charging efficiency are utilized to characterize the charging performance. We considered a scenario in which the charger energy is not maximal and the QB is not empty initially, with particular emphasis being putting on the role of different initial conditions on the charging performance. Our results showed that there exists a range of $\hat{\mu}\cdot \hat{r}_{12}$ determined by the angle between the transition dipole moments and the interatomic axis, within which the QB can be charged. Specifically, a small or large $\hat{\mu}\cdot \hat{r}_{12}$ helps to improving the charging performance. The maximal ergotropy charged on the QB and the charging efficiency decrease with the increase of the distance $r_{12}$ between the charger and the QB and vanishes or becomes infinitesimal when $r_{12}$ is large enough, whereas the corresponding charging time increases with the increase of $r_{12}$. For the small $r_{12}$ case, the maximal ergotropy charged on the QB increases with the increasing initial ergotropy in the charger, while it is weakly dependent on the residual ergotropy in the QB. This suggests that, in general, to achieve an optimal performance for this battery model, one can tune the magnitude of $\hat{\mu}\cdot \hat{r}_{12}$ such that it equals 0 or 1, to shorten the distance $r_{12}$ between the charger and the QB, and to start the charging process from a passive state of the QB. These results could provide valuable insights into the working principle of QBs and underscore the importance of considering different initial conditions when analyzing their charging performance.

\section*{ACKNOWLEDGMENTS}
This work was supported by the National Natural Science Foundation of China (Grant No. 12275212, No. T2121001, No. 92265207, and No. 92365301), Shaanxi Fundamental Science Research Project for Mathematics and Physics (Grant No. 22JSY008), the Youth Innovation Team of Shaanxi Universities, and Technology Innovation Guidance Special Fund of Shaanxi Province (Grant No. 2024QY-SZX-17).

\newcommand{\PRL}{Phys. Rev. Lett. }
\newcommand{\RMP}{Rev. Mod. Phys. }
\newcommand{\PRA}{Phys. Rev. A }
\newcommand{\PRB}{Phys. Rev. B }
\newcommand{\PRD}{Phys. Rev. D }
\newcommand{\PRE}{Phys. Rev. E }
\newcommand{\PRX}{Phys. Rev. X }
\newcommand{\APL}{Appl. Phys. Lett. }
\newcommand{\NJP}{New J. Phys. }
\newcommand{\JPA}{J. Phys. A }
\newcommand{\JPB}{J. Phys. B }
\newcommand{\PLA}{Phys. Lett. A }
\newcommand{\NP}{Nat. Phys. }
\newcommand{\NC}{Nat. Commun. }
\newcommand{\EPL}{Europhys. Lett. }
\newcommand{\AdP}{Ann. Phys. (Berlin) }
\newcommand{\AoP}{Ann. Phys. (N.Y.) }
\newcommand{\QIP}{Quantum Inf. Process. }
\newcommand{\PR}{Phys. Rep. }
\newcommand{\SR}{Sci. Rep. }
\newcommand{\JMP}{J. Math. Phys. }
\newcommand{\RPP}{Rep. Prog. Phys. }
\newcommand{\PA}{Physica A }
\newcommand{\CMP}{Commun. Math. Phys. }
\newcommand{\SCPMA}{Sci. China-Phys. Mech. Astron. }


\end{document}